\documentclass[sigconf]{acmart}
\AtBeginDocument{%
  }

\copyrightyear{2025}
\acmYear{2025}
\setcopyright{acmlicensed}\acmConference[UMAP Adjunct '25]{Adjunct Proceedings of the 33rd ACM Conference on User Modeling, Adaptation and Personalization}{June 16--19, 2025}{New York City, NY, USA}
\acmBooktitle{Adjunct Proceedings of the 33rd ACM Conference on User Modeling, Adaptation and Personalization (UMAP Adjunct '25), June 16--19, 2025, New York City, NY, USA}
\acmDOI{10.1145/3708319.3733655}
\acmISBN{979-8-4007-1399-6/2025/06}

\usepackage{float}
\usepackage{multirow}
\usepackage[skip=2pt]{caption}

\begin{document}

\title{Towards Explainable Temporal User Profiling with LLMs}

\author{Milad Sabouri}
\affiliation{%
   \institution{DePaul University}
   \city{Chicago}
   \country{USA}}
\email{msabouri@depaul.edu}

\author{Masoud Mansoury}
\affiliation{%
   \institution{Delft University of Technology}
   \city{Delft}
   \country{Netherlands}}
\email{m.mansoury@tudelft.nl}

\author{Kun Lin}
\affiliation{%
   \institution{DePaul University}
   \city{Chicago}
   \country{USA}}
\email{klin13@depaul.edu}

\author{Bamshad Mobasher}
\affiliation{%
   \institution{DePaul University}
   \city{Chicago}
   \country{USA}}
\email{mobasher@cs.depaul.edu}

\renewcommand{\shortauthors}{Sabouri et al.}

\begin{abstract}
Accurately modeling user preferences is vital not only for improving recommendation performance but also for enhancing transparency in recommender systems. Conventional user-profiling methods—such as averaging item embeddings—often overlook the evolving, nuanced nature of user interests, particularly the interplay between short-term and long-term preferences. In this work, we leverage large language models (LLMs) to generate natural language summaries of users’ interaction histories, distinguishing recent behaviors from more persistent tendencies. Our framework not only models temporal user preferences but also produces natural language profiles that can be used to explain recommendations in an interpretable manner. These textual profiles are encoded via a pre-trained model, and an attention mechanism dynamically fuses the short-term and long-term embeddings into a comprehensive user representation. Beyond boosting recommendation accuracy over multiple baselines, our approach naturally supports explainability: the interpretable text summaries and attention weights can be exposed to end users, offering insights into why specific items are suggested. Experiments on real-world datasets underscore both the performance gains and the promise of generating clearer, more transparent justifications for content-based recommendations.
\end{abstract}


% \ccsdesc[500]{Information systems~Personalization}


\maketitle

\section{Introduction}
Explainability has emerged as a critical requirement for recommender systems, enabling end users to understand—and hence trust—the reasons behind particular suggestions \cite{zhang2020explainable}. In content-based settings \cite{lops2011content}, where item metadata and textual descriptions drive user–item matching, many approaches continue to focus narrowly on maximizing predictive accuracy. They often overlook interpretability, commonly representing user preferences by simply averaging the embeddings of all items a user has interacted with \cite{polignano2021together}. While this “centric” approach is straightforward, it neglects the evolving nature of user interests and offers limited insight into \emph{why} a recommendation is made.

To address these limitations, we propose a method that not only models user preferences more accurately but also inherently supports explanation. Our framework leverages large language models (LLMs) \cite{zhao2023survey} to transform each user’s interaction history into two textual “profiles”: one highlighting short-term behaviors and another capturing stable, long-term tastes. These natural-language profiles are then encoded via a pre-trained BERT \cite{devlin-etal-2019-bert} model, resulting in two high-dimensional embeddings. An attention mechanism \cite{waswani2017attention} dynamically fuses these embeddings into a single representation that drives recommendations. Crucially, because the original short-term and long-term summaries are human-readable, they serve as transparent rationales—an essential step toward explainability. 
We validate our approach on real-world datasets across two distinct domains. In particular, these domains differ in their average user-profile size, enabling us to investigate whether an LLM-driven fusion of short-term and long-term preferences is effective in contexts with more dynamic or stable user behaviors. The textual summaries generated for each user not only support interpretability for end users but also can serve as a valuable tool for system designers, offering insight into the model’s representation of user preferences and the rationale behind specific recommendations. 

In addition, the attention mechanism applied over short-term and long-term embeddings exposes the relative influence of recent versus persistent user interests in shaping each recommendation. This dual-level transparency, combining human-readable summaries with interpretable model dynamics, can address practical demands in numerous settings where stakeholders increasingly require model outputs to be interpretable.

In this paper, we explore whether our LLM-based approach can improve ranking performance across domains characterized by different average profile sizes. We also show how the framework’s textual summaries and attention mechanism can contribute to interpretability, clarifying for end users whether a recommendation is driven more by recent consumption patterns or by stable, enduring interests. Moreover, we investigate the relative impact of each architectural component in achieving strong results.

Our contributions are threefold. First, we propose a user-profiling strategy that integrates temporal signals into semantically rich, text-based representations, naturally suited for explanation. Second, we incorporate an attention mechanism that flexibly adapts to each user’s evolving interests, providing a clear, interpretable view of how these interests guide recommendations. Finally, we present an empirical evaluation on datasets reflecting contrasting profile sizes and preference dynamics, confirming that our approach not only boosts recommendation accuracy but also furnishes valuable explanatory insights. By unifying temporal user modeling with inherent interpretability, this work sets the stage for more transparent, trustworthy content-based recommender systems.

\section{Related Works}
Explainable recommendations \cite{zhang2020explainable} have increasingly become a focal point in recommender systems research, with recent works emphasizing the need to balance transparency and accuracy. For example, \cite{chicaiza2023explainable} proposes a theoretical framework for integrating explanations into recommendation pipelines, illustrating how even simple explanatory interfaces can boost user trust. Similarly, \cite{ai2025explainable} interpretability by summarizing content and applying a linear attention mechanism, showing that clear textual justifications can coexist with competitive predictive performance. The trade-off between transparency and accuracy is examined in \cite{zhou2025based}, which leverages user reviews to produce more interpretable suggestions. Meanwhile, user studies such as \cite{chatti2022more} highlight how varying the depth of explanations can significantly impact user satisfaction, indicating that the detail level should be adaptive rather than fixed.

Recent research has also begun exploring large language models (LLMs) as a foundation for explainability. \cite{lubos2024llm} demonstrates that LLMs can generate high-quality, personalized explanations that help users better evaluate recommended items. Building on such capabilities, XRec\cite{ma2024xrec} uses a lightweight collaborative adaptor to help LLMs interpret user-item interactions, thus producing thorough, human-readable explanations. 

Meanwhile, Knowledge graph-based approaches \cite{guo2020survey} have leveraged side information to enhance transparency. Studies like \cite{shi2024llm} integrates knowledge graphs with LLM-generated reviews to highlight the reasoning paths behind recommendations, enabling interpretable item discovery even in cross-selling scenarios. Likewise, \cite{shimizu2022explainable} introduces a powerful model-intrinsic method that extends KGAT (Knowledge Graph Attention Network) \cite{wang2019kgat}. Specifically, they leverage large volumes of structured side information—including users’ demographics, item attributes, and multiple one-to-many relations—and compress it via latent-class modeling. Doing so reduces computational overhead and preserves interpretability by enabling direct visualization of which edges or neighbors drive the recommendation. This approach is particularly relevant to real-world e-commerce scenarios, in which companies store extensive multi-factor data. Their improvements highlight how “soft edges” can be used to refine attention weights and speed up training, thereby removing a key practical bottleneck of knowledge-graph-based explainable models.

Our work shares similarities with \cite{shimizu2022explainable}, as both emphasize explainability in recommendation systems enhanced by advanced neural components. However, our methodology diverges in the nature of input data utilized: whereas \cite{shimizu2022explainable} relies on side information derived from knowledge graphs, we leverage textual representations derived from user interaction histories. Specifically, we prompt large language models (LLMs) to summarize short-term versus long-term user behaviors, generating coherent and interpretable narratives suitable for direct user consumption. This integration of natural language within the recommendation pipeline effectively connects semantic insights from LLMs with temporal user behavior dynamics. Furthermore, our model explicitly differentiates between recent interests and persistent preferences, merging these aspects via an attention mechanism to produce intrinsically interpretable user profiles.

Despite substantial advances \cite{zhang2020explainable, ai2025explainable, chicaiza2023explainable, shimizu2022explainable} in generating robust and comprehensible explanations, existing methods predominantly focus either on static user profiles or strictly sequential behavioral patterns. Consequently, they often fail to account for the significant differences between transient interests and stable, long-term user preferences. While certain recommendation methods \cite{tan2016improved, kang2018self, zhu2017next} incorporate temporal dynamics, they generally do not utilize LLM-generated narratives that clarify the temporal evolution of recommendations to end users. In contrast, our method explicitly partitions user interaction histories into short-term and long-term textual summaries, subsequently encoding and integrating these summaries through an attention mechanism to support recommendation generation. Importantly, the original textual summaries remain directly accessible as built-in explanations of user interests, transparently illustrating the interplay between recent and enduring preferences without requiring additional annotations or extensive post-processing. Thus, our approach bridges the gap between advanced LLM-driven explainability and nuanced temporal modeling of user behavior.

\begin{figure*}[ht]
    \centering
    \includegraphics[width=\textwidth]{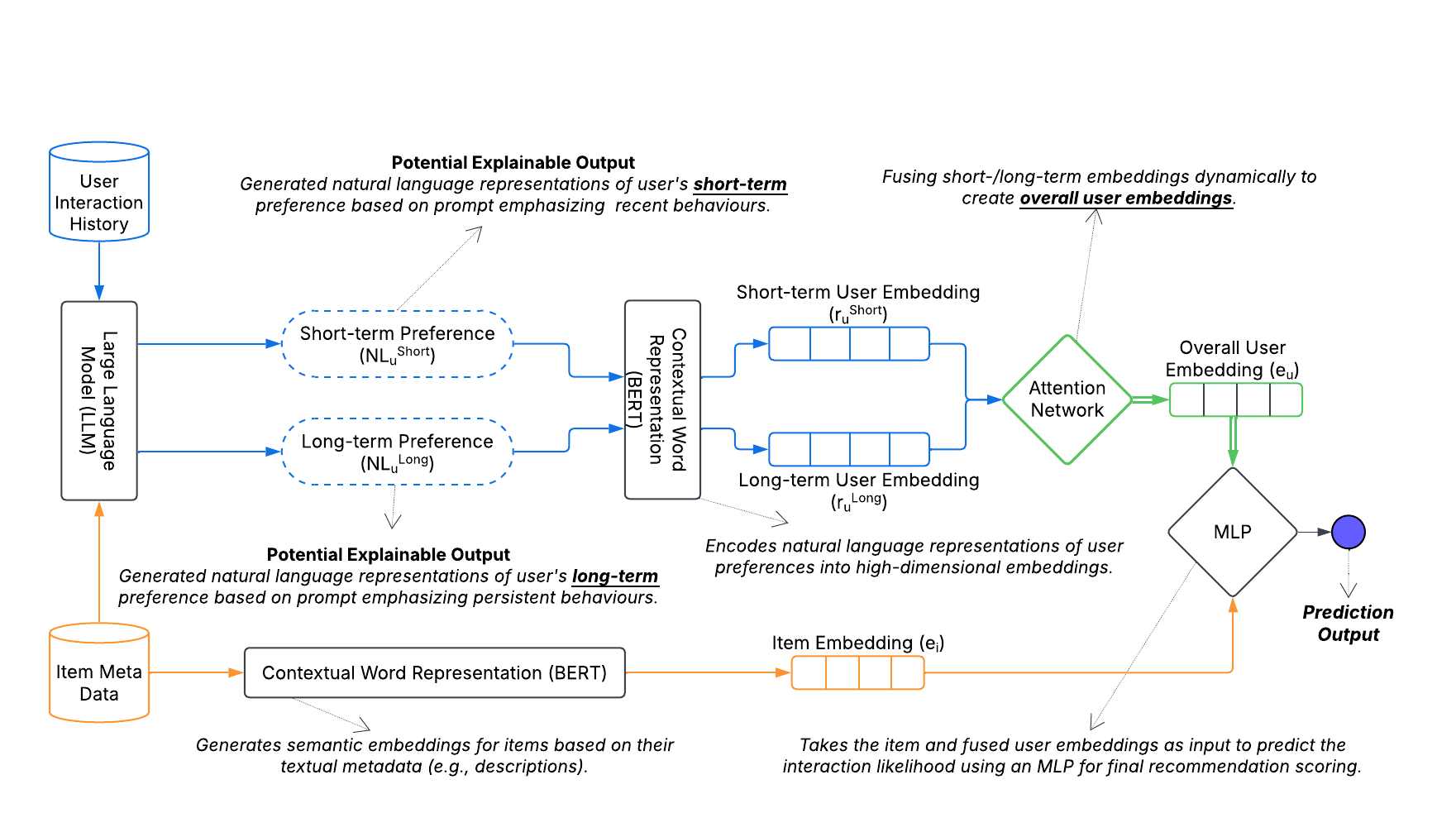}
    \vspace{-10pt}
    \caption{Proposed Architecture for LLM-Driven Temporal User Profiling}
    \label{fig:workflow}
    \vspace{-10pt}
\end{figure*}

\section{Explainable LLM-Driven Temporal User Profiling}

Let $\mathcal{U} = \{u_1, u_2, \ldots, u_{|\mathcal{U}|}\}$ denote the set of users, and $\mathcal{I} = \{i_1, i_2, \ldots\allowbreak\ ,i_{|\mathcal{I}|}\}$ denote the set of items.
Each user $u \in \mathcal{U}$ interacts with a subset of items $\mathcal{I}_u \subseteq \mathcal{I}$, where each interaction is associated with a timestamp $t_{u,i}$. Let $\mathcal{H}_u = \{(i, t_{u,i}) : i \in \mathcal{I}_u\}$ represent the interaction history of user $u$, sorted in ascending order of timestamps.

The objective is to learn a \textit{user embedding} $\mathbf{e}_u \in \mathbb{R}^d$ and an \textit{item embedding} $\mathbf{e}_i \in \mathbb{R}^d$ for each user $u$ and item $i$, such that the likelihood of user $u$ interacting with item $i$ can be predicted accurately. Specifically, we aim to estimate a function $f: \mathbb{R}^d \times \mathbb{R}^d \rightarrow [0, 1]$ that predicts the interaction probability $\hat{y}_{u,i}$ as:
\begin{equation} \label{eq:1}
    \hat{y}_{u,i} = f(\mathbf{e}_u, \mathbf{e}_i),
\end{equation}
where $\hat{y}_{u,i}$ denotes the predicted likelihood of user $u$ interacting with item $i$, and $f$ is implemented as a multi-layer perceptron (MLP). 

The proposed architecture for explainable LLM-driven temporal user profiling, illustrated in Figure~\ref{fig:workflow}, leverages temporal information and semantic representations to generate personalized \emph{and interpretable} user embeddings. The model consists of three key components: (i) Temporal User Profile Creation, (ii) Embedding Representation, and (iii) Recommendation Generation.

Beyond capturing user dynamics, each component also contributes to transparency. Our framework generates human-readable text and user-specific attention weights, both of which can be exposed to end users or analysts as intrinsic explanations of why each recommendation is made.

\subsection{User Profile Creation}

To construct a temporal-aware user profile, we generate two distinct \textbf{natural language representations} of user preferences: \textit{short-term preferences} and \textit{long-term preferences}. Let $\mathcal{H}_u = \{(i, t_{u,i}) : i \in \mathcal{I}_u\}$ denote the interaction history of user $u$, where $t_{u,i}$ represents the timestamp of interaction with item $i$. The interaction history is sorted in ascending order of timestamps to maintain chronological order.  We utilize a Large Language Model (LLM) to process the entire user interaction history twice, using distinct prompts to generate separate representations for short-term and long-term preferences:

\paragraph{Pass 1: Short-Term Profile Generation}
The complete interaction history $\mathcal{H}_u$ is provided as input to the LLM along with a task-specific prompt that instructs the model to extract the user’s \textit{short-term interests}, placing greater emphasis on the most recent interactions while still considering temporal context. The LLM generates a natural language representation of short-term preferences, denoted as:

\begin{equation}
    \text{NL}_u^{\text{short}} = \text{LLM}(\mathcal{H}_u, \text{Prompt}^{\text{short}}).
\end{equation}

\paragraph{Pass 2: Long-Term Profile Generation}
The same interaction history $\mathcal{H}_u$ is passed to the LLM again, but this time with a different prompt that instructs the model to generate a \textit{long-term preference profile} by considering the user’s entire history, capturing overarching interests and persistent behavioral patterns. The resulting representation is:

\begin{equation}
    \text{NL}_u^{\text{long}} = \text{LLM}(\mathcal{H}_u, \text{Prompt}^{\text{long}}).
\end{equation}

By explicitly utilizing distinct prompts while keeping the full interaction history intact, this approach enables the LLM to contextually differentiate between recent and persistent user preferences, addressing the limitations of traditional centric-based approaches that fail to incorporate temporal user dynamics.

Because these profiles are output as coherent text, they can be exposed directly to end users or system designers as an explanatory rationale. For example, the short-term profile might read:

\begin{quote}
\textit{``The user recently interacted with several action-thriller titles and has shown interest in fast-paced content.''}
\end{quote}

Such natural language descriptions \emph{explicitly} reveal which recent patterns are driving the recommendation. Meanwhile, the long-term profile might include broader statements about the user’s historical favorites or consistent behavioral patterns, giving a transparent view of persistent interests.

\subsection{Embedding Representation}

The generated natural language representations are encoded into high-dimensional embeddings using a pre-trained BERT model~\cite{devlin-etal-2019-bert}, resulting in short-term user embedding and long-term user embedding, as follows:
\begin{equation}\label{eq:2}
    \mathbf{r}_u^{\text{short}} = \text{BERT}(\text{NL}_u^{\text{short}}),
\end{equation}
\begin{equation}\label{eq:3}
    \mathbf{r}_u^{\text{long}} = \text{BERT}(\text{NL}_u^{\text{long}}).
\end{equation}
Here, $\mathbf{r}_u^{\text{short}} \in \mathbb{R}^d$ and $\mathbf{r}_u^{\text{long}} \in \mathbb{R}^d$ are $d$-dimensional embeddings representing the short-term and long-term preferences of user $u$, respectively. Equations~\ref{eq:2} and~\ref{eq:3} describe the encoding process in which BERT converts each natural language profile into a dense vector representation. These embeddings capture both the semantic richness of user preferences and contextual nuances from interaction histories, serving as the basis for constructing the final user profile. Similarly, item descriptions are processed by BERT to generate item embeddings, ensuring that user and item representations lie in the same embedding space.

To construct the overall user embedding, an attention mechanism~\cite{waswani2017attention} is applied to dynamically fuse the short-term and long-term user embeddings. The attention mechanism computes personalized importance weights for short-term and long-term preferences, enabling the model to adaptively emphasize recent interactions for users with rapidly changing interests or highlight stable, overarching tendencies for users with persistent behaviors. This dynamic fusion mechanism is a key differentiator from the centric approach, where user profiles are created by simply averaging item embeddings, thereby neglecting temporal nuances.

Let $\alpha_u^{\text{short}}$ and $\alpha_u^{\text{long}}$ denote the attention weights assigned to the short-term and long-term preference embeddings, respectively. The attention weights are computed as:
\begin{equation}\label{eq:4}
    \alpha_u^{\text{short}} = 
    \frac{\exp(\mathbf{W}_a \mathbf{r}_u^{\text{short}})}{\exp(\mathbf{W}_a \mathbf{r}_u^{\text{short}}) + \exp(\mathbf{W}_a \mathbf{r}_u^{\text{long}})},
\end{equation}
\begin{equation}\label{eq:5}
    \alpha_u^{\text{long}} = 1 - \alpha_u^{\text{short}},
\end{equation}
where $\mathbf{W}_a \in \mathbb{R}^{1 \times d}$ is a learnable parameter vector, and $\exp(\cdot)$ denotes the exponential function. Equation~\ref{eq:4} computes the attention weight for short-term preferences, while Equation~\ref{eq:5} ensures that the sum of these attention weights equals 1.

The overall user embedding $\mathbf{e}_u$ is then obtained as a weighted sum of the short-term and long-term embeddings:
\begin{equation}\label{eq:6}
    \mathbf{e}_u = \alpha_u^{\text{short}} \cdot \mathbf{r}_u^{\text{short}} 
    + \alpha_u^{\text{long}} \cdot \mathbf{r}_u^{\text{long}},
\end{equation}
where $\mathbf{e}_u \in \mathbb{R}^d$ represents the final user embedding. As shown in Equation~\ref{eq:6}, the attention mechanism enables the model to dynamically balance the influence of short-term and long-term preferences based on user-specific interaction patterns.

The learned attention weights $\alpha_u^{\text{short}}$ and $\alpha_u^{\text{long}}$ can be surfaced in a real system to communicate the \emph{relative importance} of the user’s recent vs.\ historical activities. For instance, if $\alpha_u^{\text{short}}$ is very high for a user, it indicates that the model believes the user’s latest interactions dominate their current recommendations—a straightforward, interpretable explanation for why certain items appear more prominently.

\subsection{Recommendation Generation}

The final user embedding $\mathbf{e}_u$, obtained from the attention mechanism, is concatenated with the corresponding item embedding $\mathbf{e}_i$ and passed through a multi-layer perceptron (MLP)~\cite{popescu2009multilayer} to predict the interaction probability of user $u$:
\begin{equation}\label{eq:7}
    \hat{y}_{u,i} = \text{MLP} \bigl([\mathbf{e}_u; \mathbf{e}_i]\bigr),
\end{equation}
where $[\mathbf{e}_u; \mathbf{e}_i] \in \mathbb{R}^{2d}$ denotes the concatenation of the user embedding $\mathbf{e}_u$ and the item embedding $\mathbf{e}_i$, and $\hat{y}_{u,i} \in [0, 1]$ represents the predicted probability that user $u$ will interact with item $i$.

As shown in Equation~\ref{eq:7}, the MLP acts as a scoring function that learns non-linear interactions between user and item embeddings. The final output is a probability indicating the likelihood of interaction. The MLP consists of multiple fully connected layers with ReLU activation functions, followed by a sigmoid output layer to ensure that the output lies in the range $[0, 1]$. The model is trained using a binary cross-entropy loss, where the target label $y_{u,i} \in \{0, 1\}$ indicates whether user $u$ interacted with item $i$:
\begin{equation}\label{eq:8}
    \mathcal{L} = - \frac{1}{|\mathcal{D}|} 
    \sum_{(u, i, y_{u,i}) \in \mathcal{D}} 
    \Bigl( y_{u,i} \log \hat{y}_{u,i} + (1 - y_{u,i}) \log (1 - \hat{y}_{u,i}) \Bigr),
\end{equation}
where $\mathcal{D}$ represents the set of all user--item interaction pairs in the training dataset. Equation~\ref{eq:8} penalizes the difference between the predicted probability $\hat{y}_{u,i}$ and the true label $y_{u,i}$, driving the model to learn accurate predictions.

% Beyond outputting $\hat{y}_{u,i}$, our framework retains textual short-term and long-term profiles, as well as attention signals, that directly illuminate \emph{why} this user--item pair scored highly. In a deployed interface, one could display:

% \begin{quote}
% \textit{``We recommend Item X because your short-term profile suggests a strong recent interest in sci-fi themes, and your long-term history shows a consistent preference for futuristic storylines.''}
% \end{quote}

% Such a statement reveals the underlying reasoning—tying the recommended item back to the user’s textual LLM-derived profile and the relative emphasis from the attention mechanism.

\subsection{Model Highlights and Strengths}

Compared to a simpler ``centric'' baseline that averages item embeddings without distinguishing short-term from long-term preferences, the present architecture offers two key benefits:

(1) \textit{Temporal Awareness:} By explicitly modeling short-term and long-term preferences and fusing them via an attention mechanism, the model better captures \emph{how} user interests shift over time, leading to improved performance.

(2) \textit{Intrinsic Explainability:} Each user’s short-term and long-term text profiles can be exposed directly, while the attention weights inform whether recent or persistent behaviors drive the recommendation. This provides a transparent rationale far more meaningful than a single averaged vector of past items.

In essence, the synergy of LLM-based text generation and attention-based fusion yields highly \emph{semantic} user embeddings that not only boost recommendation metrics (e.g., Recall@K, NDCG@K) but also enable end users and system developers to \emph{see} precisely which aspects of a user’s recent or long-running behaviors are influencing each recommendation. 

\begin{figure*}[ht]
    \centering
    \includegraphics[width=\textwidth]{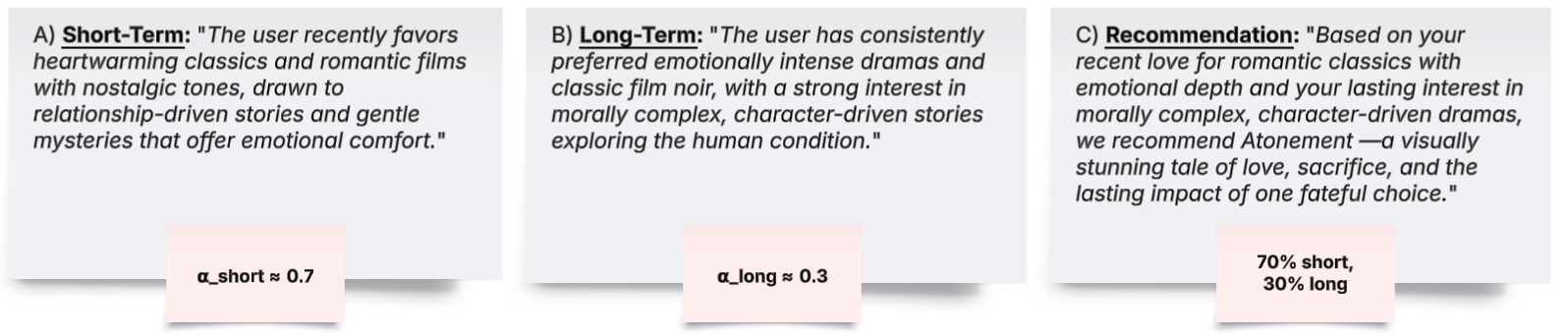}
    \vspace{-10pt}
    \caption{A conceptual illustration of our framework's potential explainability. (A) and (B) depict the short-term vs. long-term textual profiles our current approach generates and encodes. (C) shows a hypothetical extension wherein the final recommendation is explicitly justified by both sets of preferences—an aspect we plan to explore in future work.}
    \label{fig:case_study}
    \vspace{-10pt}
\end{figure*}

\section{Experiments \& Explainability Demonstration}
In this research, our primary focus is to propose an LLM-based method for creating explainable user profiles, specifically for content-based recommendation systems. The goal of these experiments is to evaluate the effectiveness of our approach in incorporating temporal dynamics, improving recommendation quality, and generating more interpretable user profiles compared to baseline models. Additionally, we showcase how our framework’s outputs can be used to offer transparent justifications for recommended items. The code, data, prompts, and experimental results are available on GitHub\footnote{https://github.com/milsab/TETUP}.

\subsection{Datasets}
We evaluated our proposed method on two public datasets drawn from Amazon Product Reviews \cite{hou2024bridging}, specifically the Movies\&TV and Video Games categories. We deliberately selected these two domains to observe our approach’s effectiveness under contrasting user-profile characteristics:

\textit{Movies\&TV:} Users in this domain often have larger and more diverse interaction histories, which are subject to frequent changes in taste (e.g., seasonal shifts, trend-driven watching habits). This makes Movies\&TV a natural testbed for short-term vs.\ long-term preference modeling. 

\textit{Video Games:} By contrast, many gamers have fewer overall interactions and more stable preferences—often remaining loyal to specific genres or franchises. This dataset thus enables us to examine whether our method’s LLM-driven, temporally aware profile construction retains benefits when user histories are relatively sparse and less variable. 

To ensure high-quality item representations, we filtered the metadata to keep only items with mostly English text and descriptions of at least 500 characters, providing enough context for the LLM to model user preferences. Both datasets provide interaction logs \textit{\emph{with timestamps}}, allowing us to incorporate temporal behavior into recommendation. Additionally, each item’s textual metadata (titles, descriptions) was leveraged to create BERT-based item embeddings, used by our method and all baselines for a fair comparison. Table \ref{tab:dataset_statistics} presents the high-level dataset details—total users, total items, and overall interaction counts. Meanwhile, Table \ref{tab:user_statistics} provides user-profile metrics such as the mean, median, and standard deviation of user interactions. Taken together, these two tables reveal the notable difference between Movies\&TV—where users typically have larger, more varied histories—and Video Games—where user profiles tend to be smaller and more stable.

\begin{table*}[t]
\centering
\small
\caption{Datasets Statistics}
\label{tab:dataset_statistics}
\begin{tabular}{lccccc}
\toprule
\textbf{Dataset} & \textbf{No. Unique Users} & \textbf{No. Unique Items} & \textbf{No. Interactions} \\
\midrule   
Movies\&TV & 10,000  & 14,420 & 202,583 \\
Games & 10,371  & 3,790 & 83,842 \\ 
\bottomrule
\end{tabular}
\end{table*}

\subsection{Baselines}
Because our work centers on user profiling for content-based recommendation, we primarily compare our proposed LLM-driven strategy against user-profiling methods commonly employed in such systems. These baselines differ in how they model user histories, incorporate (or exclude) temporal factors, and utilize textual or collaborative information. By including diverse approaches, we thoroughly test the incremental benefit of (1) large language models and (2) explicit short-term vs.\ long-term preference profiling.

\subsubsection{Centric Approach (Centric)}
As a main baseline, we use a \emph{centric} approach wherein user profiles are created by averaging all item embeddings the user has interacted with (no temporal segmentation). Other components, such as item embeddings and the final MLP architecture, remain identical to our proposed method to ensure a fair comparison. This reflects the common practice \cite{fan2023zero} of deriving user vectors by straightforward aggregation, offering a clear benchmark against which we can measure the gains from explicitly modeling recent vs.\ persistent behavior.

\subsubsection{Temporal Fusion without LLM-Based Representations (Temp-Fusion)}
This baseline represents a time-aware content-based approach that forgoes LLM-generated textual summaries. It splits a user’s interaction history into short-term and long-term segments (with dataset-specific cutoffs), encodes each segment via BERT embeddings, and applies an attention mechanism to fuse them—similar to our framework’s temporal fusion. However, rather than generating natural language profiles to be encoded by BERT, it simply aggregates the raw item embeddings from the user’s recorded interactions. By contrasting this design against our LLM-based strategy, we isolate how much semantically rich text-based representation impacts performance and explainability.

\subsubsection{Popularity-Based Recommendation (Popularity)}
This non-personalized baseline ranks items solely based on their overall popularity, measured by the frequency of interactions aggregated across all users. As a control condition, this baseline represents a minimal standard of recommendation performance that does not involve personalization or user-item modeling. Its inclusion allows us to contextualize the effectiveness of our proposed method by contrasting against a naïve strategy that ignores both user preference and content signals. Any substantial improvement over this baseline indicates the necessity of modeling beyond mere popularity trends.

\subsubsection{Matrix Factorization (MF)}
Matrix Factorization is a widely used collaborative filtering method that learns latent representations for users and items based only on observed interaction patterns. As a control condition, MF offers a direct point of comparison for evaluating our content-based approach. It captures user preferences through interaction data but lacks any understanding of item content or metadata. By including MF, we isolate the impact of incorporating content features into the recommendation process. This baseline helps clarify whether observed improvements are due to the content-aware nature of our model, rather than the presence of collaborative patterns alone—particularly in settings where user-item interactions are sparse or noisy.

Collectively, these baselines illuminate the contributions of our approach’s core components: \emph{temporal modeling}, \emph{LLM-based semantic representation}, and \emph{attention-based fusion}. By systematically comparing results, we show that our LLM-driven model, which yields both accurate predictions and transparent textual rationales, consistently outperforms less sophisticated user-profiling methods.

\begin{table}[t]
\centering
\small
\caption{User Profiles Statistics}
\label{tab:user_statistics}
\begin{tabular}{lcccc}
\toprule
\textbf{Dataset} & \textbf{Mean} & \textbf{Median} & \textbf{Mode} & \textbf{Std Dev} \\
\midrule
Movies\&TV & 11.79 & 9.00 & 6 & 9.80 \\
Games & 4.55  & 3.00 & 3 & 3.97  \\ 
\bottomrule
\end{tabular}
\end{table}

\subsection{Evaluation Methodology}
Because we aim to improve \emph{content-based user profiling}, we focus on how effectively each method learns and applies user embeddings for ranking items. Thus, the centric approach—which simply averages item embeddings over a user’s history—serves as a fundamental baseline.
Unlike sequential recommenders \cite{kang2018self, sun2019bert4rec, tan2016improved} that evaluate next-item prediction via metrics like Hit Rate or MRR, we adopt a ranking-oriented protocol. Specifically, we compute Recall@K and NDCG@K by asking each model to rank relevant (held-out) items amidst a larger candidate set. This directly tests \emph{how well the user profiles capture user intent}, which is central to content-based personalization rather than pure sequence prediction.

To incorporate temporal awareness in the training process, we sort each user’s interactions chronologically and split them into:
1) \textit{Training set:} Earliest 60\% of interactions, 2) \textit{Validation set:} The subsequent 20\% of interactions, and 3)
\textit{Test set:} Most recent 20\% of interactions.
This time-wise splitting strategy ensures that user preferences evolve naturally from training to test, allowing us to evaluate the models’ ability to predict future interactions based on historical behavior.

\subsection{Experimental Setup}
All experiments were implemented in PyTorch \cite{paszke2019pytorch} and trained using the Adam optimizer. Early stopping (patience=5) halted training if no validation improvement was seen for 5 epochs, preventing overfitting. Each model trained up to 100 epochs, with a batch size tuned among \{512, 1024, 2048\}; we present results at 2048 for best overall performance.

To encode the textual user summaries we generate for short-term and long-term profiles, we used the all-MiniLM-L6-v2 variant of SBERT \cite{reimers-2019-sentence-bert}, which yields 384-dimensional embeddings. Our method extracts short-term and long-term natural language representations of each user’s interaction history via the GPT-4o-mini model \cite{achiam2023gpt}, known for high-quality text generation. The short-term, long-term, and item embeddings all have dimension 384.

A multi-layer perceptron (hidden dimension=128, dropout=0.2) was used to predict user-item interaction probabilities. We included 5 negative samples per positive interaction. The learning rate was fixed at 1e-3, which provided stable and reliable convergence. 

The experiments were executed on an environment using four NVIDIA A100 GPUs, each with 80GB of memory, enabling efficient parallelization and acceleration of the training process. This setup, along with robust optimization techniques and careful hyperparameter tuning, ensured reliable and reproducible results across all experiments.

\begin{table*}[ht]
    \centering
    \small
    \caption{Performance Comparison of our proposed method and baselines on Movies\&TV and Video Games datasets for \(K=10\) and \(K=20\). (\textit{asterisk indicates that the improvement of the proposed method over the baseline method (Centric) is statistically significant (\(p < 0.05\))})}
    \label{tab:performance_comparison}
    \resizebox{\textwidth}{!}{ % Resizing to fit nicely
    \begin{tabular}{lllllllll}
        \toprule
        \multirow{2}{*}{\textbf{Method}} & \multicolumn{4}{c}{\textbf{Movies\&TV}}  & \multicolumn{4}{c}{\textbf{Video Games}} \\
        \cmidrule(lr){2-5} \cmidrule(lr){6-9} 
        % \textbf{Method} & \multicolumn{2}{c}{\textbf{K=10}} & \multicolumn{2}{c}{\textbf{K=20}} & \multicolumn{2}{c}{\textbf{K=10}} & \multicolumn{2}{c}{\textbf{K=20}}\\
        % \cmidrule(lr){2-3} \cmidrule(lr){4-5} \cmidrule(lr){6-7} \cmidrule(lr){8-9}
        & \textbf{Recall@10} & \textbf{NDCG@10} & \textbf{Recall@20} & \textbf{NDCG@20} & \textbf{Recall@10} & \textbf{NDCG@10} & \textbf{Recall@20} & \textbf{NDCG@20} \\
        \midrule
        (a) Centric & 0.0113 & 0.0191 & 0.0199 & 0.0269 & 0.0645 & 0.0532 & 0.0932 & 0.0649 \\
        (b) Popularity & 0.0082 & 0.0145 & 0.0133 & 0.0191 & 0.0397 & 0.0324 & 0.0706 & 0.0453 \\
        (c) MF & 0.0048 & 0.0085 & 0.0087 & 0.0124 & 0.0457 & 0.0370 & 0.0754 & 0.0491 \\
        (d) Temp-Fusion & \underline{0.0118} & \underline{0.0201} & \underline{0.0207} & \underline{0.0276} & \textbf{0.0693} & \textbf{0.0589} & \underline{0.0982} & \textbf{0.0712} \\
        (e) \textbf{LLM-Based (Proposed)}  & \textbf{0.0132}$^*$ & \textbf{0.0217}$^*$ & \textbf{0.0223}$^*$ & \textbf{0.0293}$^*$ & \underline{0.0665}$^*$ & \underline{0.0547}$^*$ & \textbf{0.1021}$^*$ & \underline{0.0683}$^*$ \\
        \midrule
        \textbf{Gain of Proposed vs. (a)} & \textbf{17\%} & \textbf{14\%} & \textbf{12\%} & \textbf{9\%} & \textbf{3\%} & \textbf{3\%} & \textbf{10\%} & \textbf{5\%} \\
        \bottomrule
    \end{tabular}
    }
\end{table*}

\subsection{Experimental Results \& Discussion}
We now evaluate the effectiveness of our LLM-based approach on two datasets: Movies\&TV and Video Games. Table \ref{tab:performance_comparison} details our method’s ranking performance for Movies\&TV and Video Games, while Table \ref{tab:user_statistics} provides a closer look at user profiles characteristics. Two observations emerge:

\textit{1) Effectiveness in Larger, Frequent-Shifting Profiles:}\\
In the Movies\&TV domain, each user’s profile tends to be bigger (mean of 11.79 interactions, with a high standard deviation of 9.80), implying diverse and more rapidly changing preferences. Table \ref{tab:performance_comparison} shows our approach significantly improves Recall@10 (by 17\%) and NDCG@10 (by 14\%) over Centric, with similarly strong gains for Recall@20 and NDCG@20 (12\% and 9\% increases, respectively). These results suggest that when users accumulate larger histories and exhibit more varied consumption patterns, short-term vs.\ long-term fusion—enriched by LLM-based text—delivers notable accuracy gains. Among the baselines, Temp-Fusion comes closest but lacks the semantic depth of our natural language profiles, reinforcing the advantage of explicit textual summaries in capturing nuanced, evolving tastes.

\textit{2) Mixed Results in Smaller, Steadier Profiles:}\\
By contrast, Video Games users’ profiles are typically shorter (mean of 4.55 interactions, median of 3.0) with a lower standard deviation (3.97), indicating more stable and less frequent content shifts. Under such conditions, our approach still achieves the highest Recall@20, affirming that even minimal LLM-derived text can help with broader item ranking. However, Temp-Fusion narrowly edges out our method on Recall@10, NDCG@10, and NDCG@20. We attribute this partly to:

\textit{Sparse Histories:} With only a few recorded interactions per user, splitting into short-term vs.\ long-term segments can dilute the available signals.

\textit{Relatively Fixed Interests:} Many gamers stay loyal to specific franchises or genres, reducing the benefit of dynamic preference modeling.

Overall, these findings highlight how dataset properties, particularly user-profile size and preference variability, impact the efficacy of our short-term–long-term LLM-driven approach. For domains like Movies\&TV, where user preferences shift more frequently and historical data is abundant, our method excels. In domains with smaller, steadier profiles, its edge is narrower, yet it still maintains the best performance in some metrics (Recall@20) and preserves an explainability advantage through textual profiles.

With regard to explainability, Figure \ref{fig:case_study} demonstrates how our framework’s short-term and long-term textual profiles contribute to explainable user modeling, while also suggesting a future extension for full user-facing recommendations. Concretely, we achieve \emph{explainability by construction} in two ways:

\begin{figure}
    \centering
    \includegraphics[width=0.45\textwidth]{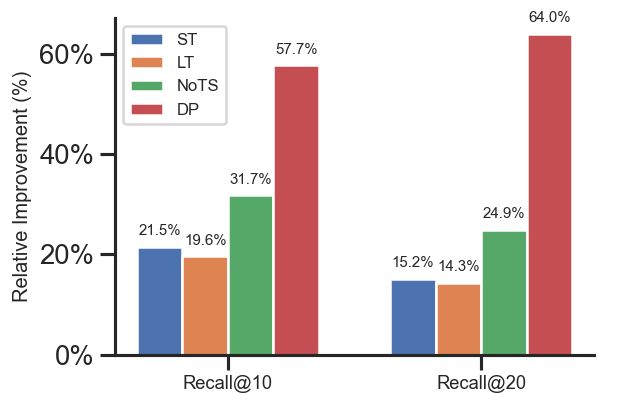}\\[1ex]
    \includegraphics[width=0.45\textwidth]{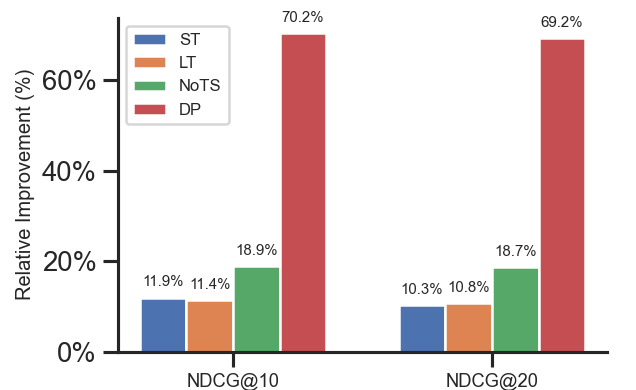}
    \caption{Relative improvement of our complete approach over four ablation variants on the Movies\&TV dataset. Bars show percentage gains in Recall@K and NDCG@K.}
    \label{fig:bar_chart}
\end{figure}

\textit{1) Textual Short-Term and Long-Term Summaries:}
Rather than producing only abstract embeddings, our model outputs natural language describing the user’s recent (Figure \ref{fig:case_study}-A) versus historical (Figure \ref{fig:case_study}-B) preferences. For instance, the short-term text might emphasize “heartwarming classics” or “romantic films with nostalgic tones,” while the long-term text highlights “emotionally intense dramas” or “morally complex, character-driven stories.” Because these summaries are easily comprehensible, users (or developers) can see exactly why certain items fit their current or enduring interests.

\textit{2) Attention-Driven Transparency:}
Each recommendation is governed by attention weights $\alpha_u^{\text{short}}$ and $\alpha_u^{\text{long}}$. In a future scenario (sketched in Figure \ref{fig:case_study}-C), if short-term interests carry 70\% of the weight, the recommendation might highlight the user’s most recent preferences. Conversely, a higher long-term weight could favor timeless favorites. This numeric breakdown is straightforward to communicate (e.g., “70\% recent tastes, 30\% long-standing preferences”), which would further reinforce user trust by clarifying why one item outperforms another.

While our current experiments focus on the textual profiles depicted in parts (A) and (B), these same signals lay the foundation for part (C): a user-facing explanation that explicitly merges short-term and long-term rationales. By unifying textual summaries with attention proportions, the final recommendation could be explained in a way that goes beyond typical opaque embeddings. Although a formal user study remains an opportunity for future work, our framework already offers transparent user profiling, which many baselines lack.

\section{Ablation Study}
We conducted an ablation study to isolate key architectural and design choices, allowing us to assess the contribution of individual components in our proposed method. The experiments include variations that use only short-term or long-term preferences, general preferences without temporal distinction, and an alternative scoring mechanism (dot product) in place of the MLP. Since the results for the Games dataset exhibited a similar pattern, we focus on the Movies dataset for clarity. Tables~\ref{tab:ablation_study_recall} and~\ref{tab:ablation_study_ndcg} show the numerical outcomes (Recall@K, NDCG@K), while Figure~\ref{fig:bar_chart} presents a bar chart illustrating how these variants compare. To structure our discussion effectively, we analyze each variant below.

\begin{table}[t]
\centering
\small
\caption{Recall@K Comparison for Ablation Variants (Movies\&TV
Dataset)}
\label{tab:ablation_study_recall}
\begin{tabular}{lll}
\hline
\textbf{Experiment} & \textbf{Recall@10} & \textbf{Recall@20} \\
\hline
\textbf{Short-Term Only (ST)} & 0.0109 & 0.0193 \\
\textbf{Long-Term Only (LT)} & 0.0110 & 0.0195 \\
\textbf{General Preferences (No TS)} & 0.0100 & 0.0178 \\
\textbf{Dot-Product Scoring (DP)} & 0.0084 & 0.0136 \\
\textbf{LLM-Based (Proposed)} & \textbf{0.0132} & \textbf{0.0223} \\
\hline
\end{tabular}
\end{table}

\begin{table}[t]
\centering
\small
\caption{NDCG@K Comparison for Ablation Variants (Movies\&TV
Dataset)}
\label{tab:ablation_study_ndcg}
\begin{tabular}{lll}
\hline
\textbf{Experiment} & \textbf{NDCG@10} & \textbf{NDCG@20} \\
\hline
\textbf{Short-Term Only (ST)} & 0.0194 & 0.0266 \\
\textbf{Long-Term Only (LT)} & 0.0195 & 0.0265 \\
\textbf{General Preferences (No TS)} & 0.0183 & 0.0247 \\
\textbf{Dot-Product Scoring (DP)} & 0.0128 & 0.0173 \\
\textbf{LLM-Based (Proposed)} & \textbf{0.0217} & \textbf{0.0293} \\
\hline
\end{tabular}
\end{table}

\subsection{Short-Term Preferences Only (ST)}
This variant exclusively focuses on short-term user interests, relying on $\text{NL}_u^{\text{short}}$ as the profile representation. We encode these textual summaries via BERT, then combine the resulting user embeddings with item embeddings in an MLP for final score computation. The goal is to isolate how well rapidly changing behaviors alone can drive recommendations, ignoring broader historical context.
As shown in Tables~\ref{tab:ablation_study_recall} and~\ref{tab:ablation_study_ndcg}, the ST variant underperforms compared to Long-Term Only (LT). This suggests that while short-term signals are helpful, they alone cannot produce a robust user profile. The gap is most pronounced at larger $K$ values. In the bar chart (Figure~\ref{fig:bar_chart}), ST’s relative improvement remains smaller than that of the full model, reinforcing the need to account for enduring user tendencies along with recent shifts.

\subsection{Long-Term Preferences Only (LT)}
Here, we rely on $\text{NL}_u^{\text{long}}$ exclusively—reflecting general, persistent user preferences—again encoded via BERT and combined with item embeddings in an MLP. This setup measures how effectively stable user interests can drive recommendations in the absence of short-term signals.
From Tables~\ref{tab:ablation_study_recall} and~\ref{tab:ablation_study_ndcg}, LT yields respectable performance, highlighting that many users show relatively consistent tastes. However, LT still lags behind our full method, as visible in Figure~\ref{fig:bar_chart}, indicating the advantage of incorporating both recent and historical behaviors rather than relying on long-term context alone.

\subsection{General Preferences without Temporal Distinction (NoTS)}
Unlike ST and LT, the “NoTS” setting collapses all user interactions into a single textual summary, dropping any notion of \emph{when} an item was consumed. By discarding short-term vs.\ long-term segmentation, we test whether explicit temporal partitioning is vital for robust user modeling.

Tables~\ref{tab:ablation_study_recall} and~\ref{tab:ablation_study_ndcg} show that NoTS outperforms the dot-product variant but remains behind ST and LT, signifying that ignoring time-based differentiation dilutes important signals. In Figure~\ref{fig:bar_chart}, NoTS consistently trails the short-term and long-term variants across the evaluated metrics. This confirms that temporal segmentation improves personalization, as recent behaviors and historical interests each contain unique information about user preferences.

\subsection{Substitute MLP with Dot Product (DP)}
This variant preserves short-term and long-term user embeddings but replaces the MLP scoring function with a simple dot product. We thus examine whether a non-linear component is necessary to fully exploit the embeddings’ expressive capacity.
The dot-product variant (DP) underperforms all others in Tables~\ref{tab:ablation_study_recall} and~\ref{tab:ablation_study_ndcg}. Figure~\ref{fig:bar_chart} further illustrates its shortcomings. Linear scoring captures fewer complex correlations, confirming that a non-linear MLP significantly reinforce ranking accuracy.

\subsection{Proposed LLM-Based Method}
Our complete approach combines short-term and long-term textual profiles, each encoded by BERT, then fused by attention before being fed into an MLP. This design accounts for both immediate user context (short-term) and stable interests (long-term), while the MLP models non-linear user--item interactions.
Across all ablation variants, our proposed method dominates in Tables~\ref{tab:ablation_study_recall} and~\ref{tab:ablation_study_ndcg}. Figure~\ref{fig:bar_chart} shows especially large gains when integrating all components—temporal separation, LLM-generated text, attention-based fusion, and an MLP. Collectively, these results confirm that each element is crucial. Short-term and long-term segmentation balances recent context with persistent tastes, LLM-based text enriches user representations, attention weights adapt to each user’s dynamic interest, and non-linear scoring provides deeper interaction modeling.
Hence, the ablation study demonstrates that integrating \emph{all} these aspects is essential for maximizing both accuracy and interpretability, validating our design choice to treat short-term vs.\ long-term signals as separately encoded textual profiles, fuse them adaptively, and employ a learnable MLP for final predictions.

\section{Conclusion}
We presented a content-based recommendation framework that integrates temporal user modeling and LLM-driven textual profiling into an inherently explainable pipeline. Rather than averaging user interactions, our approach explicitly separates short-term and long-term preferences, encodes them via BERT, and fuses the resulting embeddings with an attention mechanism. Experiments across Movies\&TV (larger, frequently shifting profiles) and Video Games (smaller, more stable profiles) confirm that time-aware, text-based user representations yield up to 17\% higher Recall@10 and 14\% higher NDCG@10 compared to a standard centric baseline.

Our method already offers built-in explainability by exposing short-term and long-term textual profiles and attention weights, clarifying whether recommendations stem from recent spikes or deeper, enduring interests. For future work, we plan to extend this design to generate fully user-facing explanations—explicitly weaving together both textual profiles and their attention ratios to explain \emph{why} a particular item is recommended.

% \bibliographystyle{ACM-Reference-Format}
% \bibliography{ref}

\end{document}